\documentclass[runningheads]{cl2emult}
\usepackage{psfig}
\usepackage{psfrag}   
\usepackage{epsf}
\usepackage{makeidx}  
\usepackage{graphicx} 
\usepackage{subeqnar} 
\usepackage{multicol} 
\usepackage{citesort}
\usepackage{amssymb}
\usepackage{amsmath}
\usepackage{latexsym}
\usepackage{cropmark} 
\usepackage{lnp}      
\makeindex            

\begin{document}

\frontmatter



\mainmatter
\setcounter{page}{203}




\title*{{\small In: T. P\"oschel and S. Luding (eds.), {\em Granular Gases}, Lecture Notes in Physics Vol. 564, Springer (Berlin, 2000), p. 203}\vspace*{0.5cm}\\ Chains of Viscoelastic Spheres}
\label{Poeschelpage}

\toctitle{Chains of viscoelastic spheres}

\titlerunning{Chains of viscoelastic spheres}

\author{Thorsten P\"oschel\inst{1}
\and Nikolai V.~Brilliantov\inst{2}$^,$\inst{1}
}

\authorrunning{T.~P\"oschel \and N.\,V.~Brilliantov}
\tocauthor{T.~P\"oschel, N.\,V.~Brilliantov}

\institute{Humboldt-Universit\"at, Institut f\"ur Physik,
  Invalidenstr. 110, D-10115 Berlin, Germany. email thorsten@physik.hu-berlin.de, http://summa.physik.hu-berlin.de/$\sim$thorsten \and Moscow State University, Physics Department, Moscow
  119899, Russia.\\ email nbrillia@physik.hu-berlin.de}

\maketitle              

\begin{abstract}
Given a chain of viscoelastic spheres with fixed masses of the first and last particles. We raise the question: How to chose the masses of the other particles of the chain to assure maximal energy transfer? The results are compared with a chain of particles for which a constant coefficient of restitution is assumed. Our simple example shows that the assumption of viscoelastic particle properties has not only important consequences for very large systems (see~\cite{BPhere}) but leads also to qualitative changes in small systems as compared with particles interacting via a constant restitution coefficient.
\end{abstract}

\section{Introduction}
\label{sec:intro}
We consider a linear chain of inelastically colliding particles of masses $m_i$, radii $R_i$ ($i=0\dots n$), with initial velocities $v_0=v>0$ and $v_i=0$ ($i=1\dots n$) at initial positions $x_i>x_j$ for $i>j$ with $x_{i+1}-x_i> R_{i+1}+R_i$. The masses of the first and last particles $m_0$ and $m_n$ are given and we address the questions: How have the masses of the particles in between to be chosen to maximize the energy transfer, i.e., to maximize the after-collisional velocity $v_n^{\prime}$ of the last particle. If $n$ is variable, how should $n$ be chosen to maximize $v_n^{\prime}$? Throughout this paper we assume that the initial distance of the particles is large enough to neglect ``multiple collisions'', i.e.,  only the first impact of each particle influences the final velocity of the $N$th sphere of the chain.

Recent investigations show that the properties of very large systems of viscoelastic particles differ significantly from those of particles interacting with constant coefficient of restitution~\cite{BPhere,NBTPDIFF,veldistr,TomThor1,NBTPage}. The system considered here may serve as an example of a {\em small} system which properties change qualitatively when the viscoelastic properties of the particles are considered.

The coefficient of restitution is defined via
$\epsilon = \left|\left({v_{i+1}^\prime}-v_{i}^\prime\right)/\left(v_{i+1}-{v_i}\right)\right|$,
which relates the relative velocity of the particles after the collision to the pre-collisional quantity. The elastic collision corresponds to $\epsilon=\mbox{const.}$ For this case, basic mechanics yields:
\begin{equation}
\label{8}
v_1^\prime=\frac{1+\epsilon}{1+\frac{m_1}{m_0}} \, v_0,~~~~v_{n}^\prime= (1+\epsilon)^n \prod_{k=0}^{n-1} \left(1+\frac{m_{k+1}}{m_k} \right)^{-1}\, v_0\,.
\end{equation}
The final velocity $v_n^\prime$ is maximized by
\begin{equation}
\label{5}
m_k=\left(\frac{m_n}{m_0}\right)^{k/n}m_0 ~,~~\mbox{yielding}~~~v_n^\prime=\left[ \frac{1+\epsilon}{1+\left(\frac{m_n}{m_0}\right)^{1/n}} \right]^{n} \,v_0\,.
\end{equation}

\begin{figure}[htbp]
\centerline{\psfig{figure=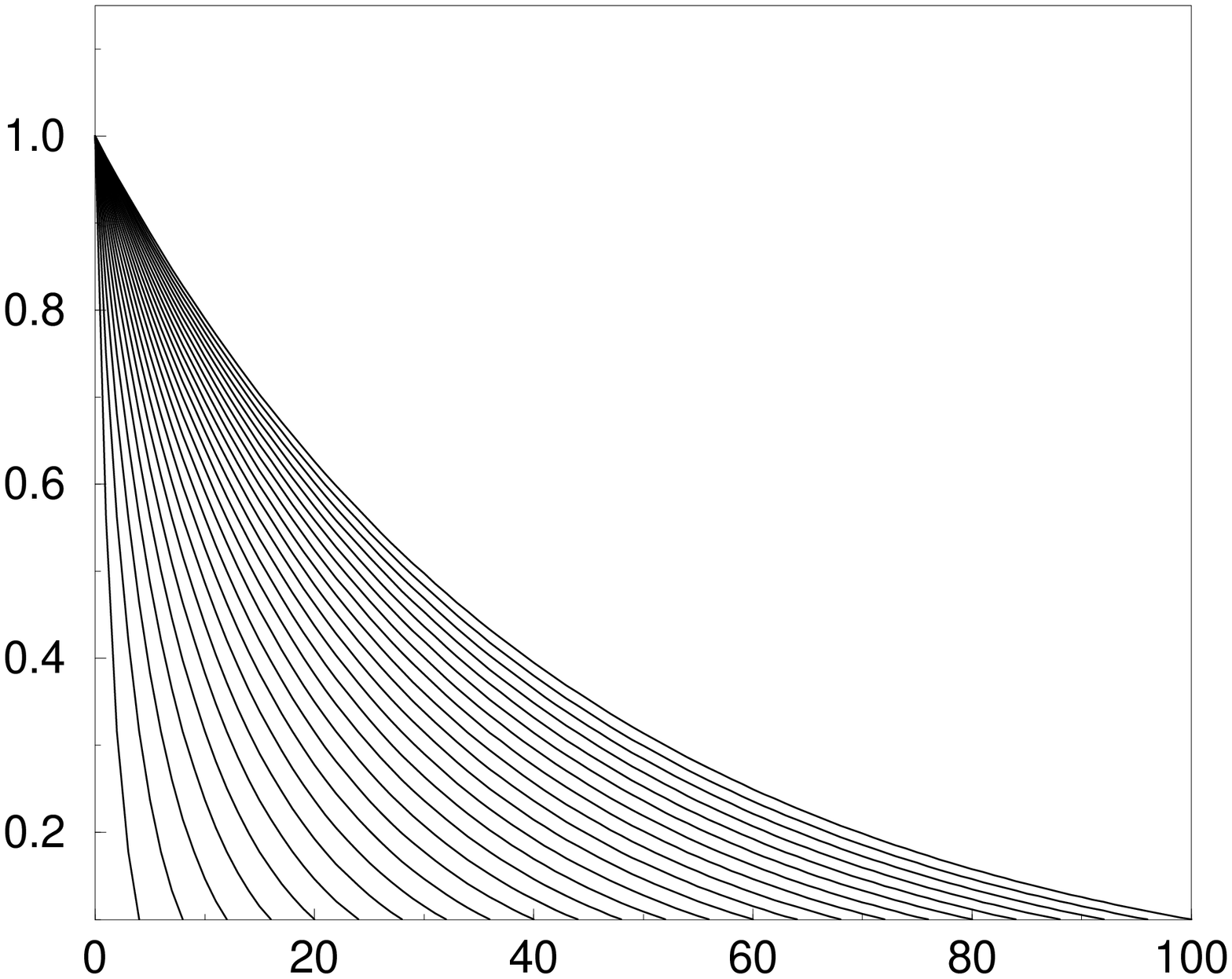,height=4.7cm,angle=270}\psfig{figure=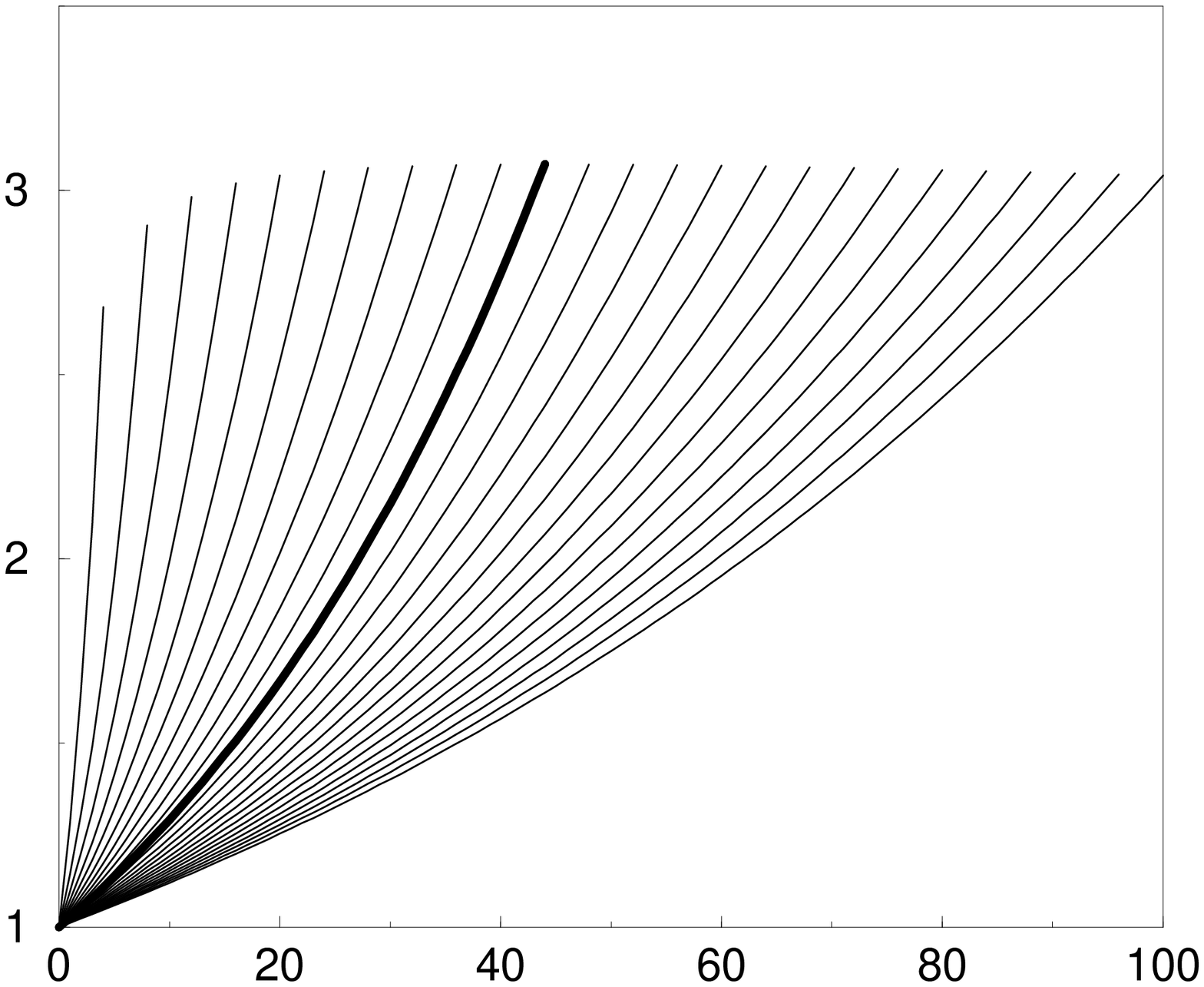,height=4.7cm,angle=270}}
    \caption{{\bf Left:} Optimal mass distribution $m_i$, $i=1\dots n$,  for the case of constant $\epsilon$. Each of the lines shows the mass $m_i$ over the index $i$ for a specified chain length $n$. The masses of the first and last particles are fixed at $m_0=1$ and $m_n=0.1$. {\bf Right:}  Velocity distributions of the particles in chains of length $n$ with the optimal mass distribution according to (\ref{5}) as a function of the chain length. The dissipative constant is $b\equiv 1-\epsilon=5\cdot 10^{-4}$. The last particle reaches its maximal velocity for chain length $n^*=44$ (bold drawn). The velocity of the first particle of the chain is  $v_0=1$.}
    \label{fig:constantMass}
  \end{figure}

The optimal mass distribution for the case of a constant coefficient of restitution $\epsilon$ does not depend on the value of $\epsilon$ and is, therefore,  the same as the optimal mass distribution in a chain of elastic particles. Figure~\ref{fig:constantMass} (left) shows the optimal mass distribution for different chain lengths $n$. The mass of the first particle is $m_0=1$ and of the last particle $m_n=0.1$. 

In contrast to the mass distribution the corresponding velocity distributions do depend on the value of the restitution coefficient $\epsilon$. Figure \ref{fig:constantMass} (right) shows the velocity distribution for $b\equiv 1-\epsilon=5\cdot 10^{-4}$. For the case of dissipative collisions the ratio $R_v=v_n^\prime/v_0$ does not monotonously increase with $n$ as for elastic particles ($\epsilon=1$), but rather it has an extremum which shifts to smaller chain lengths with increasing 
dissipative parameter  $b$. The optimal value of $n$, which maximizes $R_v$ reads
\begin{equation}
n^*=\log \left( m_0/m_n \right)/\log\left( x_0 \right) 
\label{optnconst}
\end{equation}
where $x_0$ is the solution of the equation
\begin{equation}
\label{11}
(1+x_0)=(1+\epsilon)x_0^{x_0/(1+x_0)}\,.
\end{equation}
Correspondingly, the extremal value of the $R_v$ reads
\begin{equation}
\label{12}
R_v^*=\left[ \frac{1+\epsilon}{1+x_0} \right]^{n^*}\,.
\end{equation}

\section{Chains of viscoelastic particles}

It has been shown that for colliding viscoelastic spheres the restitution coefficient depends on the masses of the colliding particles and also on their relative velocity $v_{ij}$ \cite{BSHP}. An explicit expression for the coefficient of restitution is given by the series \cite{TomThor1,Rosa} (see also \cite{BPhere})
\begin{equation}
\epsilon=1-C_1\left(\frac{3A}{2} \right)\alpha^{2/5} v_{ij}^{1/5}\!+
\!C_2 \left(\frac{3A}{2} \right)^2\alpha^{4/5} v_{ij} ^{2/5}\! \mp\! \cdots
\label{epsilon}
\end{equation}
with
\begin{equation}
\alpha= \frac{2~ Y\sqrt{R^{\,\mbox{\footnotesize eff}}}}{
3~ m^{\mbox{\footnotesize eff}}\left( 1-\nu ^2\right) }
\label{rhodef}
\end{equation}
with $Y$ and $\nu$ being the Young modulus and the Poisson ratio, respectively and $R^{\,\mbox{\footnotesize eff}}=R_iR_j/(R_i+R_j)$,
$m^{\mbox{\footnotesize eff}}=m_im_j/(m_i+m_j)$. The material constant $A$ describes the dissipative properties of the spheres (for details see~\cite{BSHP}).  The
constants $C_1=1.15344$ and $ C_2=0.79826 $ were obtained analytically
in Ref.~\cite{TomThor1} and then confirmed by numerical simulations.

In the following calculation we neglect terms ${\cal O}\left(v^{2/5}\right)$ and higher and assume for simplicity that all particles are of the 
same radius $R$, but have different masses. We abbreviate
\begin{equation}
\label{epsviab}
\epsilon=1-b\, v^{1/5}\left(m^{\mbox{\footnotesize eff}}\right)^{-2/5}~~~\mbox{with}~~~b=C_1 \left(\frac{3A}{2} \right)\left(\frac{2}{3}\frac{Y \sqrt{R/2}}{1-\nu^2}\right)^\frac25\,.
\end{equation}

\begin{figure}[htbp]
\centerline{\psfig{figure=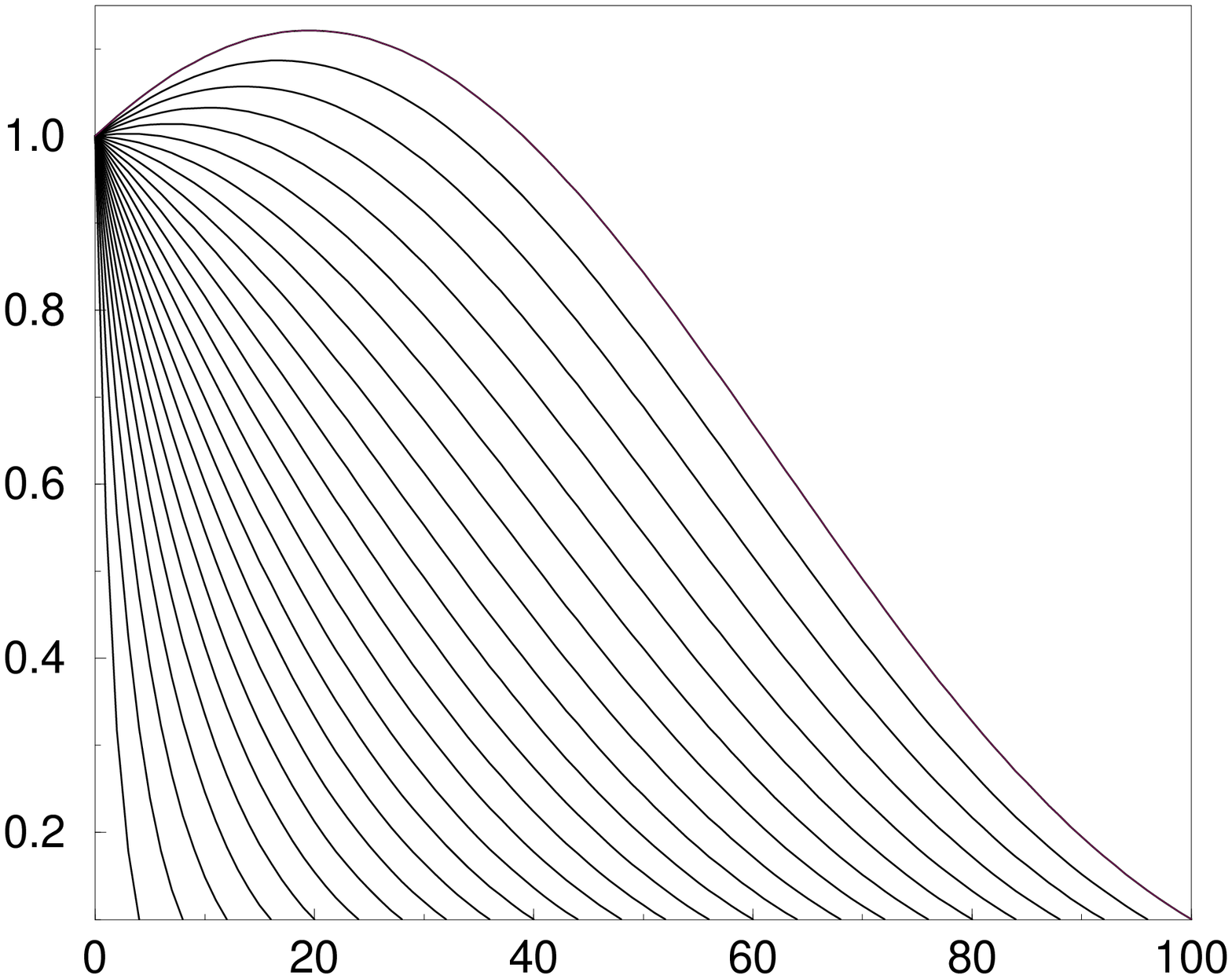,height=4.5cm,angle=270}\hspace{0.5cm}\psfig{figure=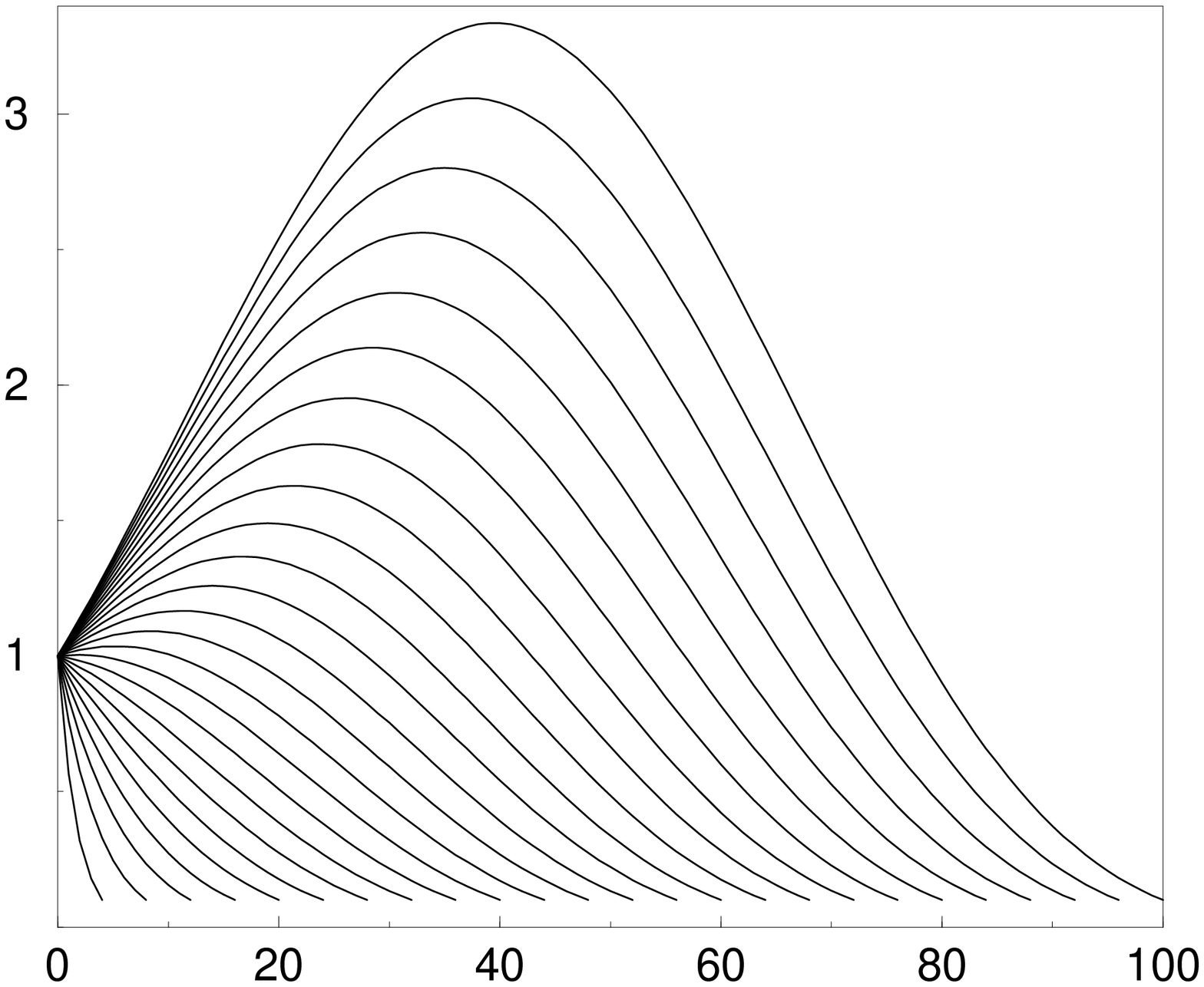,height=4.5cm,angle=270}}
\vspace{0.4cm}
    \caption{Optimal mass distribution of collision chains over the chain length $n$. The dissipative constant was $b=5\cdot 10^{-4}$ (left) and $b=2\cdot 10^{-3}$ (right).}
    \label{fig:mv22m}
  \end{figure}

Hence, for viscoelastic particles the velocities of the $k+1$-rst particle after colliding with the $k$-th reads
\begin{equation}
\label{vkk}
v^\prime_{k+1}=
\frac{2-b\left(\frac{m_{k+1}+m_k}{m_{k+1}m_k} \right)^{2/5} v_k^{1/5} }{1+\frac{m_{k+1}}{m_k}} \, v_k\,.
\end{equation}
The masses $m_k$, $k=1\dots n-1$ which maximize $v_n^\prime$ can be determined numerically and the results are shown in Fig. \ref{fig:mv22m} for two different values of the dissipative constant $b$.

\begin{figure}[htbp]
\centerline{\psfig{figure=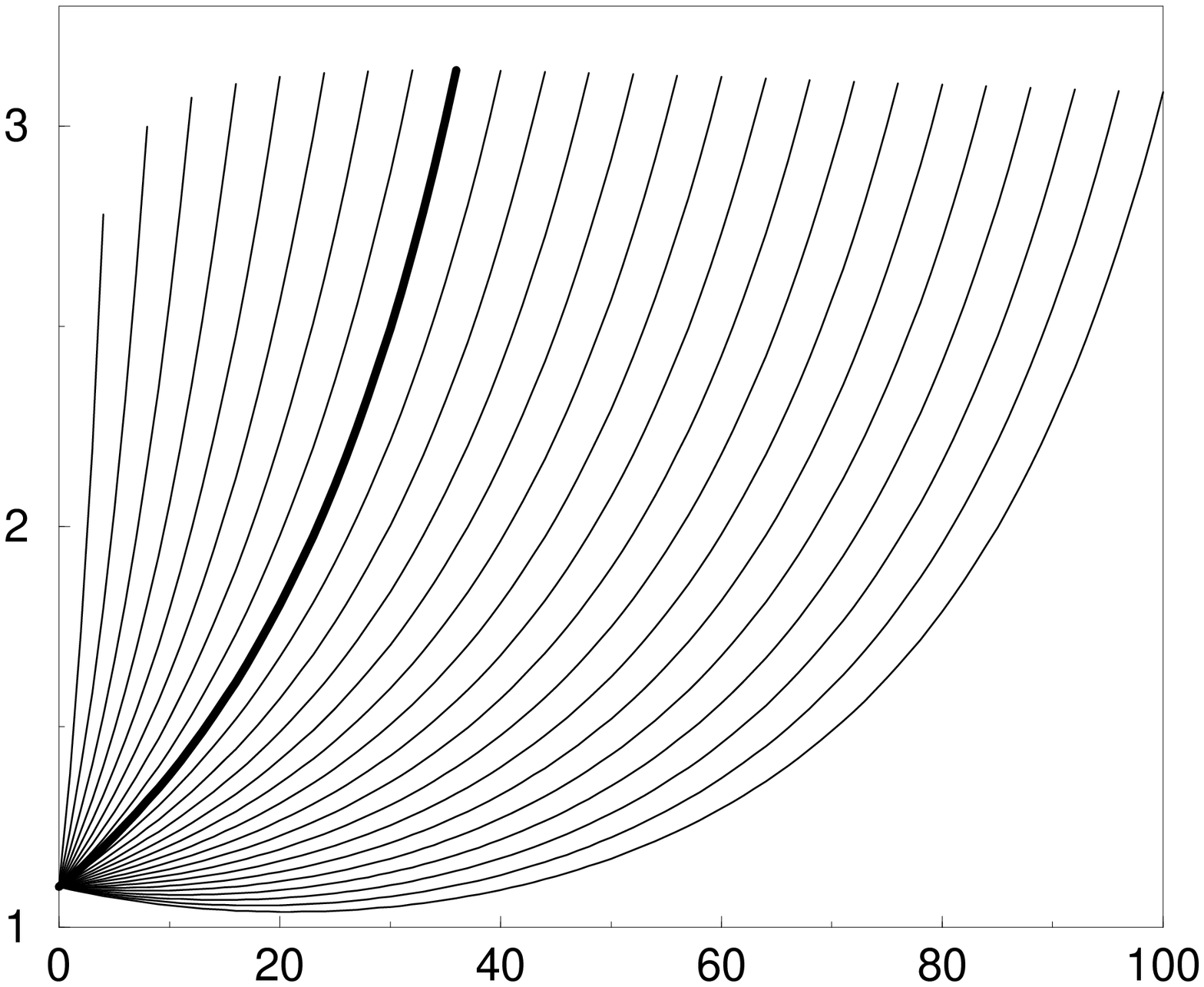,height=4.5cm,angle=270}\hspace{0.5cm}\psfig{figure=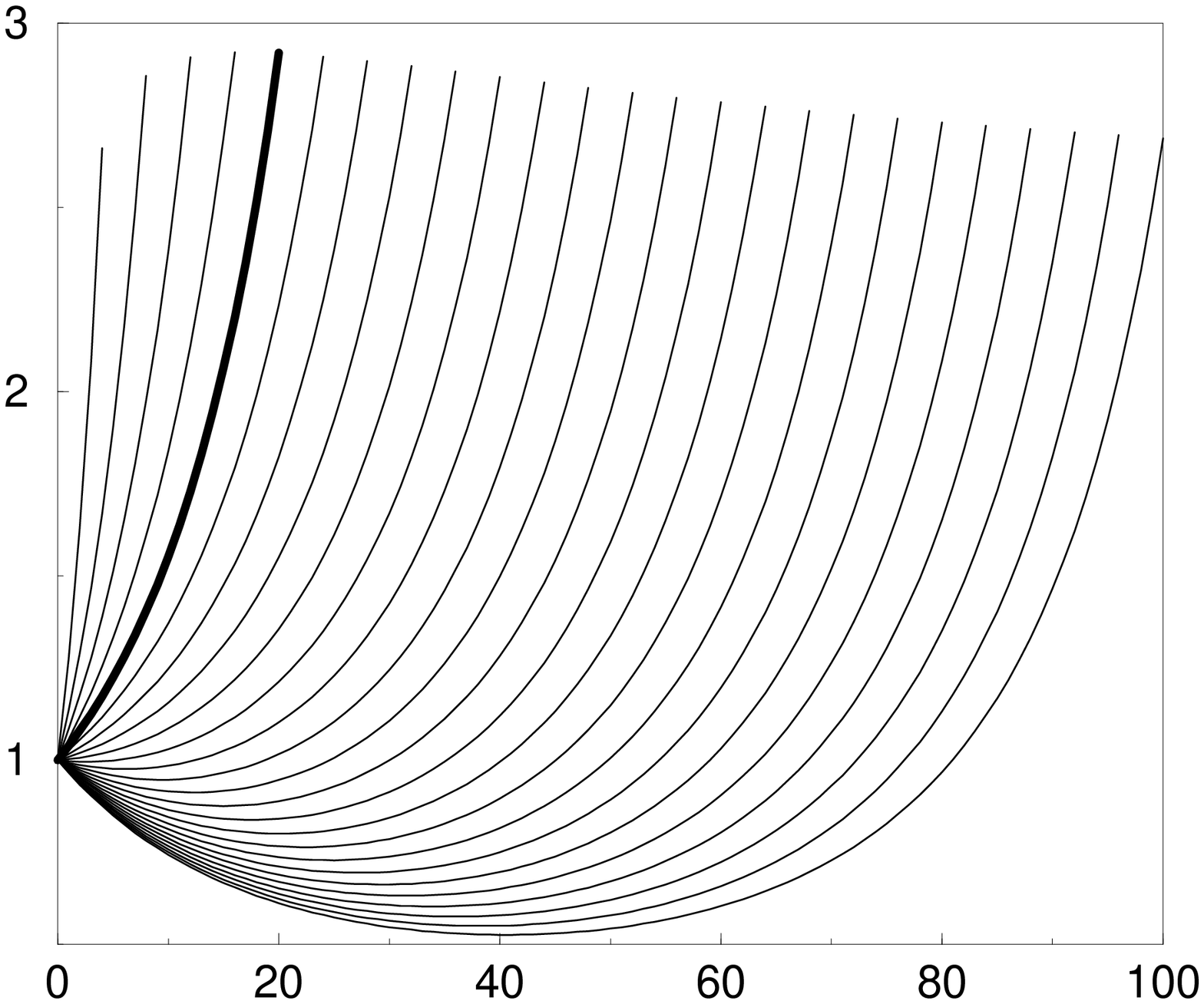,height=4.5cm,angle=270}}
\vspace{0.4cm}
    \caption{The velocities of particles in optimal chains according to Fig.~\ref{fig:mv22m}.}
    \label{fig:mv22v}
  \end{figure}

For small chain length or small $b$, respectively, the optimal mass distribution is very close to that for the elastic chain as shown in Fig.~\ref{fig:constantMass}. Again we find a monotonously decaying function for the masses. For larger chain length $n$ or larger dissipation $b$, however, the mass distribution is a non-monotonous function. The according velocities of the particles in chains of spheres of optimal masses are drawn in Fig~\ref{fig:mv22v}.

\subsection{Optimal mass-distribution}

The ``loss'' of energy, i.e., the amount of kinetic energy which is not transferred from the first particle of the chain to the last one may be subdivided into ``inertial'' and ``viscous'' losses.  Inertial losses occur due to mismatch of subsequent masses, which causes incomplete transfer of momentum even for elastic collisions if the masses differ. Viscous losses are caused by the dissipative nature of collisions. The inertial 
loss is, thus, given by the energy which remains in the $i-1$rst particle after the collision with the $i$th:
\begin{eqnarray}
\label{Ein1}
\Delta E_{in}^{(i)} &=& \frac{m_{i-1}}{2}\left(v_{i-1}^\prime\right)^2 = \frac{m_{i-1}}{2}\left(\frac{m_i-m_{i-1}}{m_i+m_{i-1}}\right)^2 v_{i-1}^2\,.
\end{eqnarray}
We describe the chain in continuum approximation $m(x)$ with $m_i\approx m_{i-1}+\frac{d m(x)}{d x}\cdot 1 $, where we assume that particles are separated on a line by unit distance. Discarding high-order mass gradients within the continuum picture $\Delta E_{in}^{(i)} \to \frac{d E_{in}}{d x} \cdot 1$  
we write for the ``line-density'' of the inertial losses
\begin{equation}
\label{Ein2}
\frac{d E_{in}}{d x} \approx \frac{\left(\frac{dm(x)}{dx}\right)^2}{8m(x)}v(x)^2\,.
\end{equation}

Viscous losses may be quantified as the difference of the kinetic energy of a particle after an {\em elastic} 
collision and  that of after a {\em dissipative} collision: 
\begin{multline}
\Delta E_{vis}^{(i)} = \left.\frac{m_{i} v_{i}^{2}}{2}\right|_{\epsilon=1}-\left.\frac{m_{i} v_{i}^2}{2}\right|_{\epsilon = \epsilon\left(v_i\right)}= \\
=\frac{m_{i}}{2}\left(\frac{2}{1+\frac{m_{i}}{m_{i-1}}}\right)^2 v_{i-1}^2 -  \frac{m_{i}}{2}\left(\frac{1+\epsilon\left(v_{i-1}\right)}{1+\frac{m_{i}}{m_{i-1}}}\right)^2 v_{i-1}^2 \\
=\frac{2 m_{i} v_{i-1}^2}{\left( 1+\frac{m_{i}}{m_{i-1}} \right)^2} 
\left\{1-\left[1-\frac{b}{2}\left(\frac{m_i+m_{i-1}}{m_i m_{i-1}}\right)^{2/5} v_{i-1}^{1/5}\right]^2 \right\}\,.
\label{viss}
\end{multline}
Expanding (\ref{viss}) up to linear order in the dissipative parameter $b$ which is assumed to be small and neglecting products of $b$ and mass gradients (which are supposed to be small too), the continuum transition of Eq.~(\ref{viss}) yields
\begin{equation}
\frac{d E_{vis}}{d x} \approx \frac{b}{2^{3/5}} m^{3/5} v^{11/5}.
\end{equation}
Thus, the total energy loss in the entire chain reads
\begin{equation}
\label{Etot1}
  E_{tot} = \int\limits_0^n \left[
\frac{m_x^2}{8m}v^2 + \frac{b}{2^{3/5}} m^{3/5} v^{11/5} \right] dx= \int\limits_0^n \left[ \frac{m_x^2}{8m^2} + \frac{b}{2^{3/5}} 
\frac{1}{m^{1/2}}\right] dx\,.
\end{equation}
with $m_x \equiv dm/dx$. For the second part of Eq. (\ref{Etot1}) we assume in zero-order approximation the ``ideal chain Ansatz'' for the velocity distribution $v(x)$, which refers to the velocity distribution $v(x)$ in an
idealized chain, where the kinetic energy completely transforms through the chain, i.e., where $\frac12 m(x)v^2(x)= {\rm const}=\frac12 m_0v_0^2$. With $m_0=1$, $v_0=1$, i.e., $v(x)=1/\sqrt{m(x)}$, the right hand side of Eq. (\ref{Etot1}) follows.

\begin{figure}[htbp]
\centerline{\psfig{figure=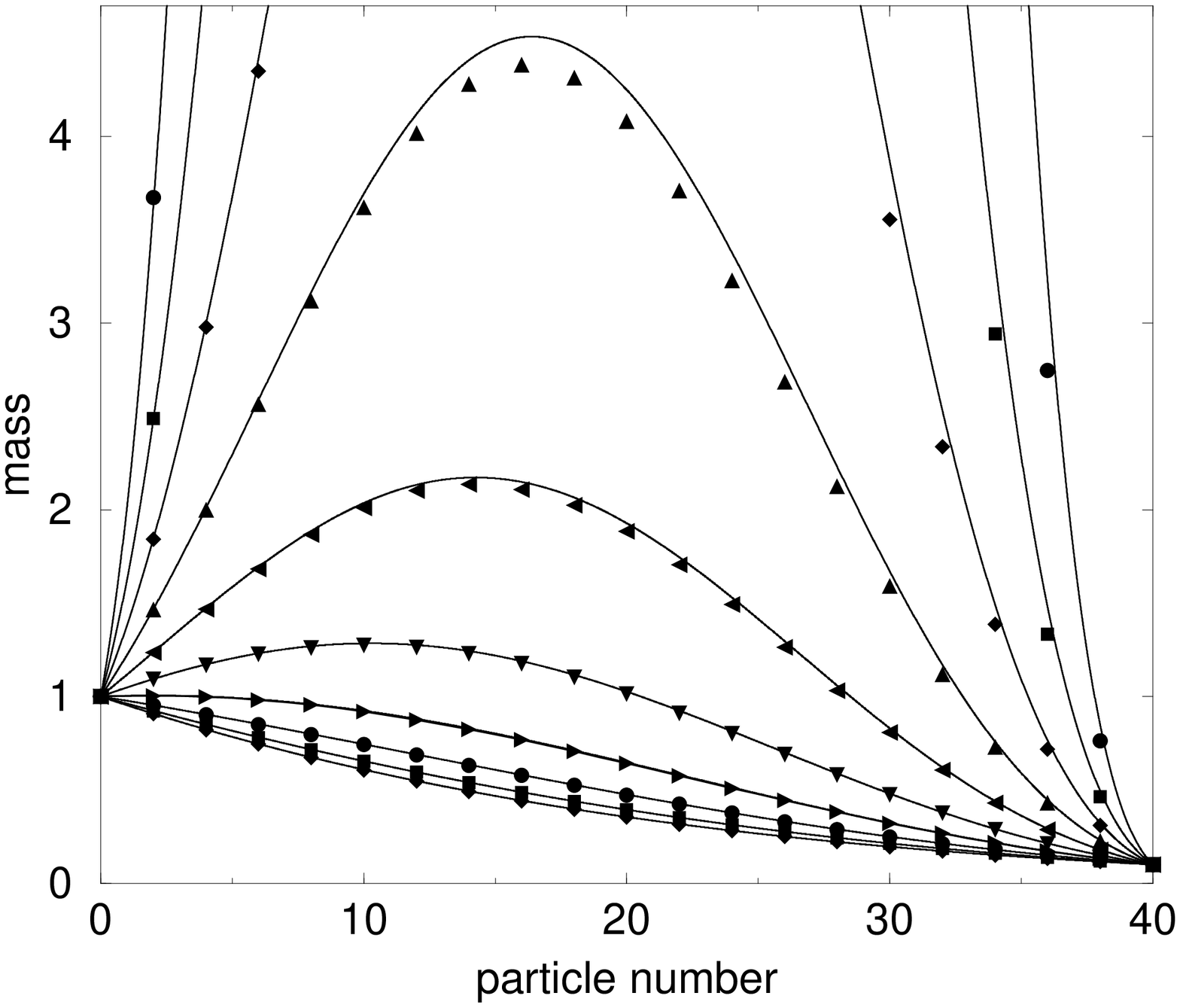,height=4.8cm,angle=270}\psfig{figure=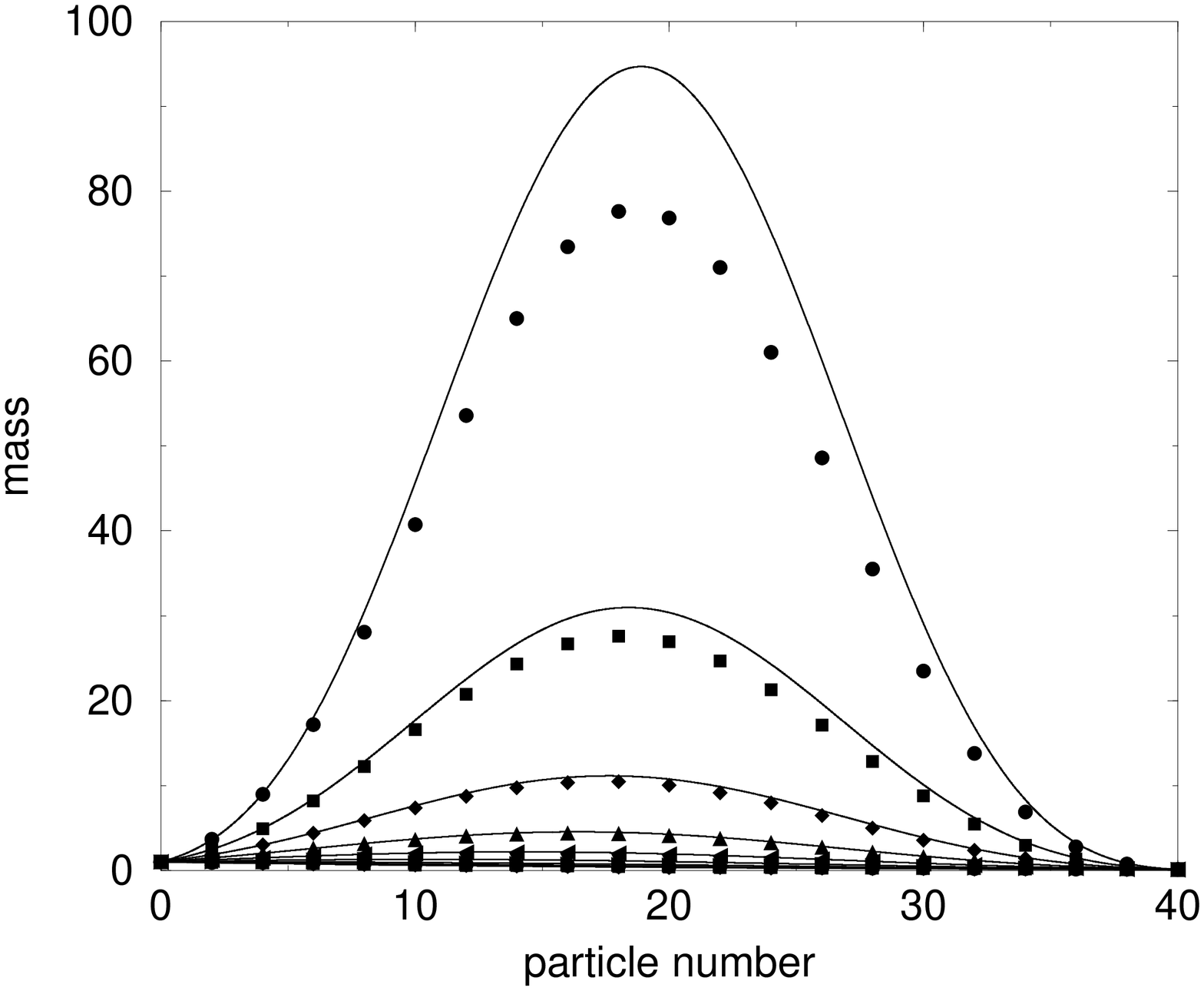,height=4.8cm,angle=270}}
\psfull
    \caption{{\bf Left:} Mass distribution of the optimal chain of length $n=40$
    for different values of the dissipative parameter $b$. Lines: numerical solution of Eq. (\ref{eqy}), 
    Points: discrete numerical optimization (from top to bottom: $\bullet:~b=0.128$,
    $\blacksquare:~b=0.064$, $\blacklozenge:~b=0.032$,
    $\blacktriangle:~ b=0.016$, $\blacktriangleleft:~b=0.008$,
    $\blacktriangledown:~b=0.004$, $\blacktriangleright:~b=0.002, $
    etc.). {\bf Right:} Same data and symbols as left but plotted in larger scale. }
    \label{fig:mAnalNumA}
  \end{figure}

The mass distribution which minimizes $E_{tot}$ satisfies the
Euler-equation applied to the integrand in (\ref{Etot1}):
\begin{equation}
\label{eq:Euler}
  \frac{d}{dx}\frac{2m_x}{8m^2} - 
\frac{\partial}{\partial m} \left[ \frac{m_x^2}{8m^2} + \frac{b}{2^{3/5}}
 \frac{1}{m^{1/2}} \right] = 0
\end{equation}
which leads to an equation for the mass distribution of
the optimal chain, written for $y(x) \equiv 1/m(x)$:
\begin{equation}
\label{eqy}
  \frac{d^2 y}{dx^2} - \frac{1}{y}\left(\frac{dy}{dx}\right)^2 -
  2^{2/5}b y^{3/2} = 0\,.
\end{equation}
Figure~\ref{fig:mAnalNumA} (lines) shows the numerical solution of (\ref{eqy}), i.e., the optimal mass distribution, for a
chain of length $n=40$ for different damping parameters $b$. The points show the results of a discrete numerical optimization of the
full chain problem  (see Eq.~(\ref{vkk})) applying a steepest descent method to optimize the masses $m_k$ of all particles. For small dissipation $b$ both results agree.

For larger values of $b$ the solution of Eq.~(\ref{eqy}) deviates from the discrete
optimization which is understandable since in our approximation we assumed the gradients of the mass distribution to be small which is violated for larger $b$. While the absolute values of masses deviates from the discrete calculation, Eq.~(\ref{eqy}) still predicts well the position of the maximum of $m(x)$.

\begin{figure}[htbp]
\centerline{\psfig{figure=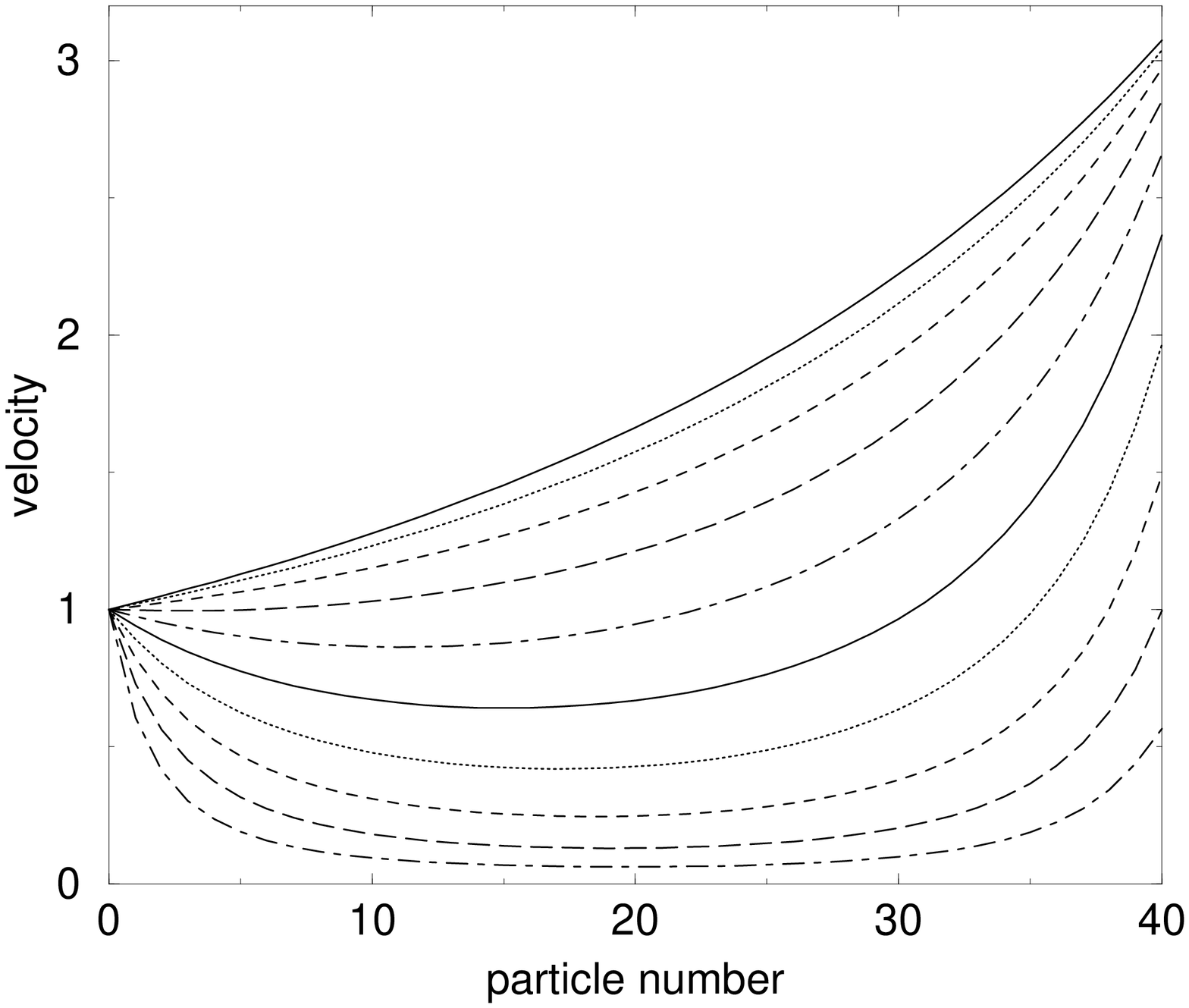,height=4.1cm,angle=270}\hspace{0.8cm}\psfig{figure=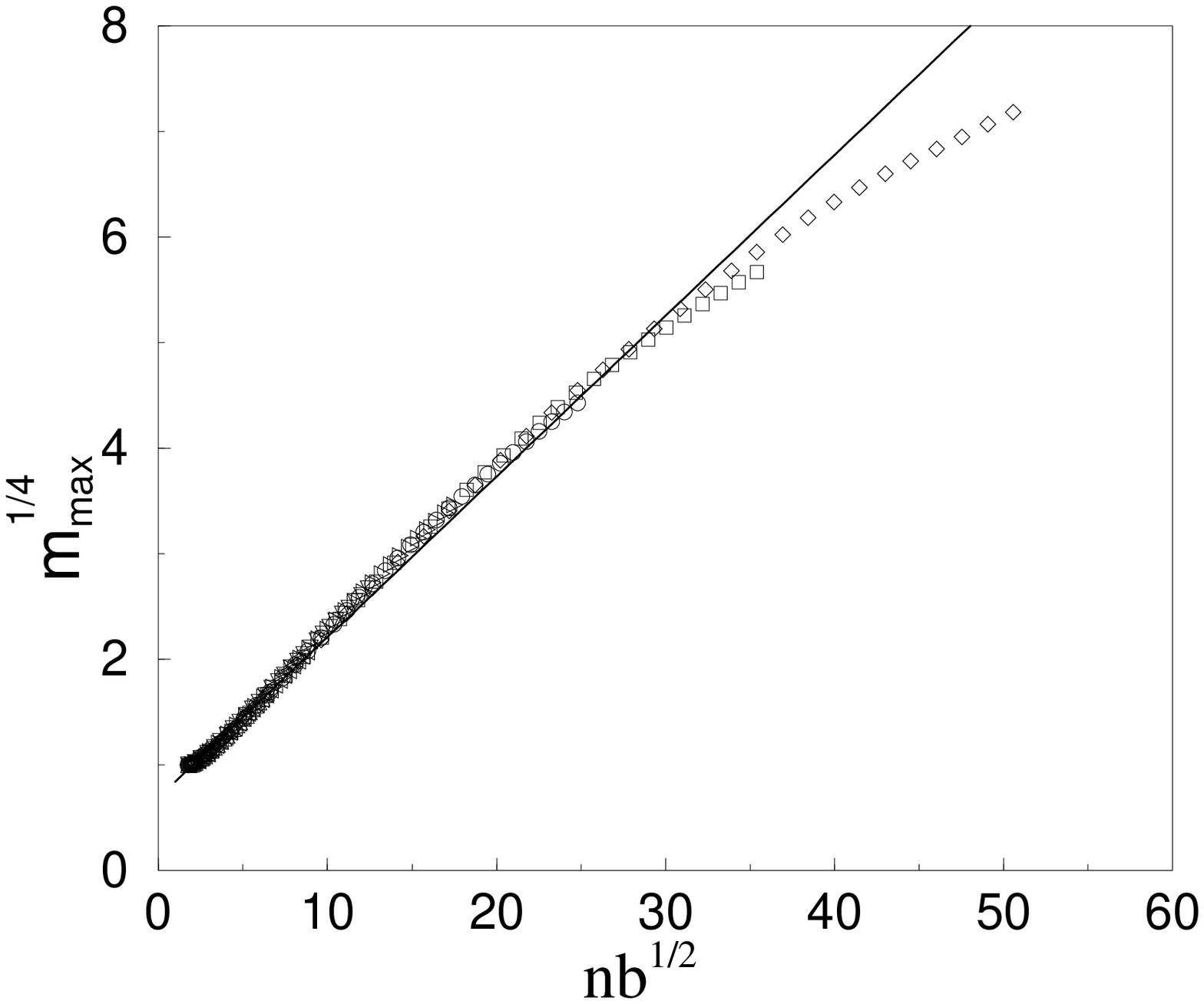,height=4.1cm,angle=0}}
\psfull
\vspace{0.8cm}
    \caption{{\bf Left:} The velocity distribution along the optimal chain shown in 
    Fig.~\ref{fig:mAnalNumA} Lines from top to bottom: $b=2.5\cdot 10^{-4}$,
    $5\cdot 10^{-4}$, $0.001$, $0.002$, $0.004$, $0.008$, $0.016$,
    $0.032$, $0.064$, $0.128$. }
    \label{fig:vepsonv40}
    \caption{{\bf Right:}  The mass of the heaviest sphere $m^*$ in an optimal chain
    depends on the dissipative parameter $b$ and on the chain length
    $n$. In the figure we plotted $\left(m^*\right)^{1/4}$ over $n\sqrt{b}$ for about
    3000 different combinations of $b$ and $n$ ($n=2\dots 300$,
    $b=0.0001\dots 0.256$) including all data presented in
    Figs. \ref{fig:mv22m}, \ref{fig:mAnalNumA}. Without any adjustable parameters the data
    from the numerical optimization of chains agrees well with the
    analytical expression Eq. (\ref{eq:maxmass}).}
    \label{fig:maxmass}
  \end{figure}

Figure \ref{fig:vepsonv40} displays the corresponding distribution of
velocities for the optimal chains shown in
Fig. \ref{fig:mAnalNumA}. According to the maximum in the mass
distribution, the velocity distribution reveals for larger $b$ a
pronounced minimum.

One can give a simple physical explanation of the appearance of a maximum in
the mass distribution (and the corresponding minimum in the velocity
distribution): As it is seen from Eq. (\ref{epsviab}) the restitution
coefficient increases with decreasing impact velocity and increasing
masses of colliding particles; this reduces the viscous losses. Thus,
slowing down particles, by increasing their masses in the inner part
of the chain, leads to decrease of the viscous losses of the energy
transfer.  The larger the masses in the middle and the smaller their
velocities, the less energy is lost due to dissipation. On the other
hand, since the masses $m_0$ and $m_n$ are fixed, very large masses in the
middle of the chain will cause large mass mismatch of the subsequent
masses and, thus, large inertial losses [see Eq. (\ref{Ein1})].  The
optimal mass distribution, minimizing the {\em total} losses,
compromises (dictated by $b$) between these two opposite tendencies.
For the case of a constant coefficient of restitution the relative
part of the kinetic energy, which is lost due to dissipation does not
depend on the impact velocity. This means that only minimization of
the inertial losses, caused by mass gradient, may play a role in
the optimization of the mass distribution. Thus, only a monotonous mass
distribution with minimal mass gradients along the chain may be
observed as an optimal one for the case of the constant restitution
coefficient.

\subsection{The maximum of the optimal mass-distribution}

The mass $m^*$ of the heaviest sphere in the optimal mass distribution can be expressed as a function of the chain
length $n$ and the dissipative parameter $b$. With the term $y^{-1}\left( dy/dx \right)^2$ discarded, Eq.~(\ref{eqy}) describes formally 
scattering of a particle of unit mass by the potential
\begin{equation}
U(y)=-\frac12 d\,b y^{5/2}~~~~~\mbox{with}~~~~~d \equiv \frac{4}{5} 2^{2/5}.
\end{equation} 
Formally changing notations $x \to t$ (``time'') to emphasize the
mechanical analogy, we write the equation of motion 
\begin{equation}
  \label{y(t)}
\ddot{y}=-\frac{dU}{dy}\,,~~~y_0=y(t=0)=\frac{1}{m_0}\,,~~~y_n=y(t=n)=\frac{1}{m_n}\,.
\end{equation}
Here we consider the case of mass distributions
having a maximum; the generalization, however, is
straightforward. Hence,
\begin{equation}
  \label{ydot}
\frac12 \dot{y}^2+U(y)={\rm const}=U(y^*)
\end{equation}
where $y^*$ is the turning point in the scattering problem, i.e., the
point where the particle's ``velocity'' $\dot{y}$ is zero (this
corresponds to the point $m^*$ of the mass distribution in the initial
problem). The ``particle'' reaches this point at ``time'' $t^*$, i.e.,
\begin{equation}
  \label{ydot1}
\dot{y}^2=d\, b \left(y^{5/2}-y^{*\, 5/2} \right)\,.
\end{equation}
Solving Eq. (\ref{ydot1}) with respect to $\dot{y}$ 
yields
\begin{equation}
  \label{ydotsolv}
\frac{dy}{dt}=
\pm \, \sqrt{d\,b} \, y^{*\, 5/4} \sqrt{\left(y/y^* \right)^{5/2}-1}\,.
\end{equation}
Integration over ``time'' from $t=0$ to $t=n$ in Eq. (\ref{ydotsolv}), therefore, leads to (with correct choice of signs)
\begin{equation}
  \label{integovert}
\frac{y^{*\, -\frac54}}{\sqrt{d\,b}}
\left[\int_{y^*}^{y_0} \frac{dy}{\sqrt{\left( \frac{y}{y^*}\right)^{\frac52 }-1}}
+\int_{y^*}^{y_n} \frac{dy}{\sqrt{\left( \frac{y}{y^*}\right)^{\frac52 }-1}}
 \right]=n\,.
\end{equation}
Using the substitute $z=(y^*/y)^{5/2}$, the integrals in Eq. (\ref{integovert}) may be recasted into the form
\begin{equation}
  \label{inttrans}
\frac25\, y^* \int\limits_{\left(\frac{y^*}{y_k} \right)^{\frac52}}^{1}
z^{-\frac{9}{10}}\left(1-z \right)^{-\frac12}dz = {\cal B} \left(\frac{1}{10}, \frac12 \right) -
{\cal B} \left[\frac{1}{10}, \frac12,\,  \left(\frac{y^*}{y_k} 
\right)^{\frac52} \right]
\end{equation}
with $k=0$ for the first integral in the the left-hand side of
Eq. (\ref{integovert}) and with $k=n$ for the second integral. 
 ${\cal B}(x,y)$ is the Beta-function and ${\cal B}(x,y,\,a)$ is
the incomplete Beta-function (which has an upper limit $a$ instead of
$1$ in its integral representation). If we assume the pronounced
maximum in the optimal mass-distribution, so that $a \equiv
\left(y^*/y_k \right)^{5/2} =\left(m_k/m^* \right)^{5/2}$ is small,
one can approximate the incomplete Beta-function as
\begin{equation}
  \label{Betaapprox}
{\cal B} \left(\frac{1}{10}, \frac12, \,  a \right) \equiv 
\int_0^a z^{-\frac{9}{10}} \left(1-z \right)^{-\frac12} dz 
\approx \int_0^a z^{-\frac{9}{10}} = 10\, a\,.
\end{equation}
With the use of Eqs. (\ref{inttrans})
 and
(\ref{Betaapprox}), Eq. (\ref{integovert}) reads
\begin{equation}
  \label{nandBeta}
\frac{2 y^{*\,-\frac14}}{5\sqrt{d\,b}}
\left\{2 {\cal B} \left(\frac{1}{10}, \frac12 \right)-
10\left[ \left(\frac{y^*}{y_0} \right)^{\frac14}+ \left(\frac{y^*}{y_n} 
\right)^{\frac14}\right]\right\}=n\,.
\end{equation}
In the original variables, which refer to the mass distribution, we
obtain a {\em scaling} relation, connecting the heaviest mass $m^*$,
the chain length $n$ and the dissipative parameter $b$:
\begin{equation}
  \label{eq:maxmass}
  \left(m^*\right)^{1/4} = p\, \sqrt{b} \, n + q
\end{equation}
with the constants 
\begin{equation}
\label{pq}
p \equiv \frac{5\sqrt{d}}{4{\cal B} \left(\frac{1}{10}, \frac12
\right)} 
~~~~~~~q \equiv \frac{5}{{\cal B}
\left(\frac{1}{10}, \frac12 \right)}
\left[ m_0^{\frac14}+m_n^{\frac14} \right]\,.
\end{equation}
So far we considered the solution of the variational Eq. (\ref{eqy})
with the term $y^{-1}\left(dy/dx \right)^2$ discarded. For this case
the constant $d$, which has been given above reads: $d=\frac45\,
2^{2/5}$. It may be shown, however, that perturbative (thus
approximate) account of this omitted term leads to an equation of the
same form as Eq. (\ref{ydotsolv}), but with the {\em renormalized}
coefficient $d \to \frac95\,d = \frac{36}{25}\,2^{2/5}$; the details 
are given in~\cite{BPchain}. Using numerical values for ${\cal B} \left(\frac{1}{10},
\frac12 \right)$ (see \cite{GradshteinRyzhik}), the renormalized coefficient
$d$ and $m_0=1$, $m_n=0.1$ for the first and the last masses of the
chain, yields for $p$ and $q$:

\begin{equation}
\label{pqval}
p=0.15217 \qquad q=0.68989
\end{equation}

In Fig. \ref{fig:maxmass} we compare the analytical relation
Eq.~(\ref{eq:maxmass}) with the constants given in Eq.~(\ref{pqval}) with numerical results for $m^*$. The numerical data follow from the numerical optimization of the mass distribution for
different chain length and different dissipative constants, including all data given in Figs.~\ref{fig:mv22m}-\ref{fig:vepsonv40}. As one can
see from Fig.\ref{fig:maxmass}, the results of the analytical theory
and of the numerical optimization agree well, except for large
dissipation values. We would like to stress that no fitting
parameters have been used.

\section{Conclusion}

We investigated analytically and numerically the transmission of
kinetic energy through one-dimensional chains of inelastically
colliding spheres. For constant restitution coefficient, $\epsilon=\mbox{const.}$, the distribution
of the masses which leads to optimal energy transfer, is an exponentially decreasing
function which is independent on $\epsilon$, i.e., it is the same as for elastic particles with $\epsilon=1$. 

For viscoelastic particles where $\epsilon$ depends on the impact velocity, the optimal mass
distribution is not necessarily a monotonous function, but depending on the chain length $n$ and on the
material parameters of the spheres it may reveal a pronounced maximum. 

We develop a theory which describes the total energy losses along the chain, so that the optimal mass distribution,
minimizing the losses, may be obtained as a solution of a variational
equation. We derived an expression relating the heaviest mass in the chain to the chain length and the dissipation constant. Having no fitting parameters,
it is in good agreement with the numerical data.

It has been demonstrated before that for the case of
``thermodynamically-large'' granular systems the impact-velocity
dependence of the restitution coefficient leads to qualitatively different behavior
as compared to systems with $\epsilon=\mbox{const.}$ (e.g.~\cite{Luding,Luding1,TomThor1,NBTPDIFF,veldistr,NBTPage,BPhere,Spahn,Spahn1,Spahn2}).  
Our system demonstrates that the velocity dependence of the
restitution coefficient leads to qualitative modifications in small and simple systems too. 
Therefore, in general, we believe that the assumption of a constant
coefficient of restitution is an approximation which justification
cannot be assumed {\em \'a priori} but has to be checked for each
particular application.

\end{document}